\documentstyle{article}
\newcount\Mac  \Mac=0 
\newcommand{\pha}{\phi_1}
\newcommand{\phit}{\tilde{\phi}}
\newcommand{\rt}{\tilde{r}}
\newcommand{\Vt}{\tilde{V}}
\newcommand{\Rt}{{\cal R}}
\newcommand{\At}{\tilde{A}}
\newcommand{\St}{\tilde{S}}
\newcommand{\Mc}{{\cal M}}
\newcommand{\Det}{{\rm Det}}
\newcommand{\phb}{\phi_2}

\newcommand{\lx}{\lambda}
\newcommand{\Lx}{\Lambda}

\newcommand{\kx}{\kappa}

\newcommand{\be}{\begin{equation}}
\newcommand{\ee}{\end{equation}}
\newcommand{\een}{\end{subequations}}
\newcommand{\ben}{\begin{subequations}}
\newcommand{\beq}{\begin{eqnarray}}
\newcommand{\eeq}{\end{eqnarray}}

\def \lta {\mathrel{\vcenter
     {\hbox{$<$}\nointerlineskip\hbox{$\sim$}}}}
\def \gta {\mathrel{\vcenter
     {\hbox{$>$}\nointerlineskip\hbox{$\sim$}}}}

\def\Red{}
\def\Black{}
\def\Blue{}

\newcommand{\lascia}[1]{}
\def\puttag(#1,#2)#3{\put(#1,#2){\makebox(0,0){\rm\Blue #3\Black}}}

\def\circa#1{\,\raise.3ex\hbox{$#1$\kern-.75em\lower1ex\hbox{$\sim$}}\,}

\def\putps(#1,#2)(#3,#4)#5#6{\ifnum\Mac=1 \put(#1,#2){\special{picture #5}}
\else  \put(#3,#4){\includegraphics{#6}} \fi}
\def\Tr{\mathop{\rm Tr}}

\def\One{\hbox{1\kern-.24em I}}
\newcommand{\phibounce}{\phi_{\rm b}}

\makeatletter
\def\art{\@ifnextchar[{\eart}{\oart}}
\def\eart[#1]#2#3#4#5#6{{\rm #2}, {\e, #3 \bf #4} {\rm (#6) #5} ({\em #1})}
\def\hepart[#1]#2{{\rm #2, \em#1}}
\newcommand{\oart}[5]{{\rm #1}, {\em #2 \bf #3} {\rm (#5) #4}}
\newcounter{alphaequation}[equation]
\def\thealphaequation{\theequation\hbox to
0.6em{\hfil\alph{alphaequation}\hfil}}
\def\eqnsystem#1{
\def\@eqnnum{{\rm (\thealphaequation)}}
\def\@@eqncr{\let\@tempa\relax \ifcase\@eqcnt \def\@tempa{& & &} \or
  \def\@tempa{& &}\or \def\@tempa{&}\fi\@tempa
  \if@eqnsw\@eqnnum\refstepcounter{alphaequation}\fi
\global\@eqnswtrue\global\@eqcnt=0\cr}
\refstepcounter{equation} \let\@currentlabel\theequation \def\@tempb{#1}
\ifx\@tempb\empty\else\label{#1}\fi
\refstepcounter{alphaequation}
\let\@currentlabel\thealphaequation
\global\@eqnswtrue\global\@eqcnt=0 \tabskip\@centering\let\\=\@eqncr
$$\halign to \displaywidth\bgroup \@eqnsel\hskip\@centering
$\displaystyle\tabskip\z@{##}$&\global\@eqcnt\@ne
\hskip2\arraycolsep\hfil${##}$\hfil& \global\@eqcnt\tw@\hskip2\arraycolsep
$\displaystyle\tabskip\z@{##}$\hfil
\tabskip\@centering&\llap{##}\tabskip\z@\cr}

\def\endeqnsystem{\@@eqncr\egroup$$\global\@ignoretrue} \makeatother

\renewcommand{\theequation}{\arabic{equation}}

\begin{document}
\twocolumn[
\centerline{\bf April 1999 \hfill     SNS-PH/99-5}
\centerline{\bf hep-ph/9904357 \hfill IFUP--TH/99--16} \vspace{1cm}
\centerline{\Large\bf\Red Bubble-Nucleation Rates for Cosmological Phase Transitions}
\bigskip\bigskip\Black
\centerline{\large\bf Alessandro Strumia} \vspace{0.3cm}
\centerline{\em Dipartimento di Fisica, Universit\`a di Pisa}
\centerline{\em e INFN, sezione di Pisa,  I-56126 Pisa, Italia}\vspace{0.3cm}
\centerline{and}\bigskip
\centerline{\large\bf Nikolaos Tetradis}\vspace{0.3cm}
\centerline{\em Scuola Normale Superiore}
\centerline{\em Piazza dei Cavalieri 7, I-56126 Pisa, Italia}
\bigskip\bigskip\Blue
\centerline{\large\bf Abstract}
\begin{quote}\large\indent
We estimate bubble-nucleation rates for cosmological phase transitions.
We concentrate on the evaluation of the pre-exponential factor, for 
which we give approximate analytical expressions. Our approach relies
on the use of a real coarse-grained potential.
We show how the coarse-graining scale can be determined in the studies of 
high-temperature phase transitions. We discuss 
the metastability bound on the Higgs-boson mass and the 
electroweak phase transition. 
We find that the saddle-point approximation is reliable in the
first case and breaks down in the second case.
\end{quote}\Black
\vspace{1cm}]

\paragraph{Introduction:} The estimates of bubble-nucleation ra\-tes for
cosmological first-order phase transitions are carried out 
within Langer's theory of homogeneous nucleation~\cite{langer}, 
applied to relativistic field theory in refs.~\cite{coleman}.
The nucleation rate is exponentially suppressed by the 
action (free energy rescaled by the temperature) of the
critical bubble, 
a saddle point of the free energy of the system.
Significant contributions to the nucleation rate 
may arise from higher orders in a systematic expansion around
this saddle point. The first correction has the form of a pre-exponential
factor that involves fluctuation determinants around the 
saddle-point configuration and the false vacuum. 
The evaluation of this factor is a difficult
problem at the conceptual and technical level, as crucial issues associated
with the convexity of the potential, the divergences of the fluctuation 
determinants 
and the double-counting of the effect of fluctuations 
must be resolved. Several approaches have been proposed in order to address
these issues~\cite{ewein,schmidt}.

In a series of recent works~\cite{first}--\cite{fourth}, following 
the proposal of refs.~\cite{bubble1}, we 
developed a consistent approach, based on the effective average action
$\Gamma_k$~\cite{averact} that
can be interpreted as 
a coarse-grained free energy.
Fluctuations with 
characteristic momenta larger than a coarse-graining scale
($q^2 \gta k^2$) are integrated out
and their effect is incorporated in 
$\Gamma_k$. 
In the limit $k \to 0$,
$\Gamma_k$ becomes equal to the effective action.
The $k$ dependence 
of $\Gamma_k$ is described by an exact flow equation~\cite{exact}.
This flow equation can be translated into evolution equations
for functions appearing in a derivative expansion of
the action~\cite{indices}. An approximation that is sufficient 
in most cases  
takes into account the effective average potential
$U_k$ and a standard kinetic term and neglects higher derivative
terms in the action.  
The bare theory is defined
at some high scale $\Lx$ that can be identified with the ultraviolet 
cutoff. 
It is, however, more convenient to 
choose a starting scale $k_0$ below the temperature $T$,
where the effective average action of
a $(3+1)$-dimensional theory at non-zero temperature
can be described in terms of an effective  
three-dimensional action at zero temperature~\cite{trans,me}.

In ref.~\cite{first} we computed the form of $U_k$ at scales $k\leq k_0$ for 
a theory of one scalar field by
integrating its evolution equation, starting with
an initial potential $U_{k_0}$ with two minima separated by a barrier.
$U_k$ is real and non-convex for non-zero $k$, and 
approaches convexity only in the limit $k\to 0$.
The nucleation rate must be computed for $k$ larger than the scale $k_f$
for which the negative curvature at the top
of the barrier becomes approximately equal to 
$-k_f^2$~\cite{convex1,convex2}. 
For $k<k_f$ the form of the potential
is affected by field configurations that interpolate between
the two minima. 
Also, for $k \gta k_f$ the typical length scale of a thick-wall critical
bubble is $\gta  1/k$. 
We performed the calculation of the nucleation rate for a range of scales
above and near $k_f$.
In our approach 
the pre-exponential factor is calculated with an 
ultraviolet cutoff of order $k$, as the effect of fluctuations
with $q^2 \gta k^2$ is already incorporated in $U_k$.
The saddle-point configuration 
has an action $S_k$ with a significant $k$ dependence. 
For strongly first-order phase transitions, the nucleation
rate $I = A_k \exp(-S_k)$
is dominated by the exponential suppression.
The main role of the prefactor $A_k$, which is also $k$ dependent,  
is to remove the scale dependence from the total nucleation rate.
For progressively more weakly first-order phase transitions,
the difference between 
$S_k$ and $\ln ( A_k/k^4_f )$ diminishes and 
contributions from higher orders in
the expansion around the saddle point become important.
At the same time, a significant $k$ dependence of the
predicted nucleation rate develops. 
The prefactor always enhances the total nucleation rate and
can compensate the exponential suppression for weakly first-order 
phase transitions.
This indicates that there is a limit to the validity of Langer's picture 
of homogeneous nucleation, set by the requirement of convergence of the
expansion around the saddle-point~\cite{second}. 

First-order phase transitions
in two-scalar models were studied in ref.~\cite{third}, where
the applicability of homogeneous nucleation theory to
radiatively-induced first-order phase transitions was tested. 
The prefactor tends to suppress the total nucleation rate
in this case.
It was found that the expansion around the saddle point
is not convergent for such phase transitions,
for which fluctuations are so strong that they generate a new minimum
in the potential. 
This indicates that
estimates of bubble-nucleation rates for the electroweak phase transition
that are based only on the saddle-point action may be very misleading. 
The reliability of our approach was reconfirmed through the study
of (2+1)-dimensional theories at non-zero temperature~\cite{fourth}
and the comparison with results from lattice simulations.

In this paper we present a simplified description of our approach that
permits quick tests of the applicability of nucleation theory in a
variety of problems. 
We give approximate expressions
for the evaluation of the pre-exponential factor in the nucleation rate:
\begin{itemize}
\item Eq.~(\ref{iter}) allows a quick determination of the 
coarse-grained
potential in terms of the standard perturbative
potential.
\item Eq.~(\ref{apprpref}) allows an approximate evaluation of the prefactor
in terms of the critical-bubble profile.
\end{itemize}
We check their validity 
by comparing them with numerical calculations of the fluctuation determinants.
Finally, we estimate the importance of the prefactor in the
cases of the bound on the Higgs-boson
mass from vacuum stability and the electroweak phase transition.

\paragraph{Simple approximation for the coarse-grained potential:}
We study the potential of a scalar field $\phi$ coupled to several 
other fields with zero expectation value. 
We consider the theory at energy scales below the temperature $T$,
so that a $(3+1)$-dimensional theory at non-zero temperature
can be described in terms of an effective  
three-dimensional one at zero temperature.
The correspondence 
between the quantities we use and the ones of the four-dimensional
theory is given by 
\beq
\phi =\frac{\phi_4}{\sqrt{T}},\qquad
U(\phi)=\frac{U_4( \phi_{4},T )}{T}.
\label{fivethree} \eeq
The evolution equation for the coarse-grained potential $U_k(\phi)$
takes the form~\cite{exact,indices,first,second,third}
\be
\frac{\partial}{\partial k^2} U_k(\phi) =
-\frac{1}{8 \pi} {\rm Tr} \sqrt{k^2 + \Mc^2_k(\phi)}, 
\label{evpot} \ee
where $\Mc^2_k(\phi) $ is the 
mass matrix of all the fields whose
mass depends on $k$ and the expectation value of $\phi$.
In order to derive the above expression we assumed that the various
fields have standard kinetic terms and neglected higher-derivative
interactions in the effective average action. We also used a
mass-like infrared cutoff $\sim k^2$ in order to eliminate 
contributions to $U_k$ from modes
with characteristic momenta $q^2 \lta k^2$.

The first step of an iterative solution of eq.~(\ref{evpot}) gives~\cite{iterative}
$$U_k^{(1)}(\phi) 
= U_{k_0}(\phi) +
\frac{1}{12\pi}\times$$
\be
\times \Tr \left\{
\left[ k_0^2 + \Mc^2_{k_1}(\phi) \right]^{\frac{3}{2}}
-\left[ k^2 + \Mc^2_{k_1}(\phi) \right]^{\frac{3}{2}}  \right\}.
\label{iter} \ee
The scale $k_1$ can be chosen 
arbitrarily between $k_0$ and $k$.
For $k=0$ and $k_1=k_0$, eq.~(\ref{iter}) is a regularized one-loop
approximation to the effective
potential. For $k_1=k > 0$, it takes the form of 
a ``mass-gap'' equation for the coarse-grained potential. It is the
latter form that we shall find more useful.

For practical applications of our method the values of
the scales $k_0$ and $k_f$ must be specified. 
For studies of high-temperature field theories the 
natural choice of $k_0$ is $k_0=T$. For $k\lta k_0$ the
system can be described in terms of an effective three-dimensional
theory at zero temperature~\cite{trans,me}. 
The form of $U_{k_0}(\phi)$ depends on 
the zero-temperature potential and the thermal
fluctuations with $q^2 > T^2$. It can be determined 
by integrating the
evolution equations for the (3+1)-dimensional theory at zero and
non-zero temperature, starting from a scale $\Lx \gg k_0$,
with the bare action as initial condition~\cite{trans}.  
In refs.~\cite{trans,me} it was checked 
that eq.~(\ref{iter}) with $k=0$ and $k_1=k_0$ reproduces
the one-loop approximation to the high-temperature effective potential.
This fact suggests a simpler procedure that was employed with
success in ref.~\cite{fourth}. 
One can demand that $U^{(1)}_k(\phi)$ is equal, at $k=0$,
to the perturbative
result (or the solution to a ``mass-gap'' equation).
It is easy then to extract $U_{k_0}(\phi)$ from eq.~(\ref{iter}).
In this way no explicit reference to the zero-temperature potential
and the thermal modes with $q^2 > T^2$ is necessary. 
Notice that the imaginary part
of the standard perturbative potential
matches with the one of our approximate solution 
$U^{(1)}_0(\phi)$ of eq.~(\ref{iter}).

The choice of $k_f$ is dictated by the form of the matrix $\Mc^2_{k_1}(\phi)$
in eq.~(\ref{iter}).
The eigenvalue $\Mc^2_{1k_1}(\phi)$ that
corresponds to the field that varies along the profile of the bubble 
is equal to the second derivative of the potential. This means that
$U_k^{(1)}(\phi)$
would become complex in the non-convex regions for
$k <k_f$, with
$k^2_f = \max|U_{k_1}''{}^{(1)}(\phi)|$.
This reproduces the well-known pathology of perturbation theory.
However, the exact solution $U_k(\phi)$
of eq.~(\ref{evpot}) does not have
imaginary parts. On the contrary, after $k$ becomes almost equal 
(within a few percent) to
${\rm max}|U_k''(\phi)|$ in the non-convex regions,
the negative curvature of the potential 
goes to zero $\sim -k^2$~\cite{convex2,analytical}, 
resulting in the Maxwell construction. This behaviour is
related to the renormalization of the potential by configurations in
the path integral that interpolate between its minima~\cite{convex1}.
As this mechanism is intrinsically non-perturbative, it is not
captured by the approximate solution of eq.~(\ref{iter}).
As was demonstrated in refs.~\cite{first}--\cite{bubble1}, a
consistent description of tunnelling can be obtained by computing the 
nucleation rate at a coarse-graining scale sufficiently large,
so that the potential is not affected by interpolating
configurations and the ``approach to convexity'' has not set in.
For our approximate discussion, the solution of 
eq.~(\ref{iter}) can be employed and
the nucleation rate can be computed above and close to $k_f$, with
$k^2_f = \max|U_{k_1}''{}^{(1)}(\phi) |$. If the expansion
around the saddle point is convergent, the value of the bubble-nucleation
rate has a weak dependence on $k$. 

\paragraph{Simple approximation for the nucleation rate:}
The nucleation rate is given by 
$$I=A_{k} \exp({-S_k})$$
where
$$A_{k}= \frac{E_0}{2\pi}\left(\frac{S_k}{2\pi}\right)^{3/2}
\bigg|
\frac{\Det'\left[-\partial^2+\Mc^2_{k}(\phibounce(r)) \right]}
{\Det \left[ -\partial^2+k^2 +\Mc^2_{k}(\phibounce(r))\right]}\times$$
\beq\times
\frac{\Det\left[-\partial^2+k^2+\Mc^2_{k}(0) \right]}
{\Det\left[-\partial^2+\Mc^2_{k}(0)\right]}
\bigg|^{-1/2}
\label{rrate} \eeq
Here $\phibounce(r)$ is the profile of the spherically-symmetric
saddle point, $S_k$ its action computed through the potential
$U_k(\phi)$,
and $\phi = 0$ corresponds to the false vacuum. 
The prime in the fluctuation determinant around
the saddle point denotes that the 3 zero eigenvalues 
of the operator $-\partial^2+\Mc^2_{k}(\phibounce(r))$, corresponding to
displacements of the critical bubble, 
have been removed. $E_0$ is the square root of
the absolute value of the unique negative eigenvalue.
For theories with continuous internal symmetries, the corresponding
zero eigenvalues are removed from the determinants as well and the nucleation
rate is multiplied by an appropriate factor
\cite{buch}.
The above form of $A_k$ guarantees that only 
modes with characteristic momenta $q^2 \lta k^2$ contribute to the
nucleation rate. We emphasize that the
use of an ultraviolet cutoff in eq.~(\ref{rrate}) that matches the
infrared cutoff procedure in the derivation of eq.~(\ref{evpot})
is crucial for the consistency of our approach. In both cases, mass-like
cutoffs have been used.

The prefactor $A_k$ can be expressed as a product of terms
$A_{ik}$ corresponding to eigenvalues $\Mc^2_{ik}$
of the mass matrix in orthogonal field directions. 
Also, using the spherical symmetry of the saddle-point configuration,
one can express each of $A_{ik}$ as a product of contributions $c_{i\ell}$
with given quantum number $\ell$~\cite{first,third}.
For large $\ell$, first-order perturbation theory in 
$W_{ik}(r)\equiv \Mc^2_{ik}(\phi_b(r))-\Mc^2_{ik}(0)$
gives~\cite{first,third}
\beq
c_{i\ell} \to 1 + \frac{D_i}{\ell^2}+ {\cal O}(\frac{1}{\ell^4}) 
\label{asympt} \eeq
with
$$
D_i = -\frac{1}{4} k^2 \int_0^\infty 
r^3\, W_{ik}(r)\,dr.$$
The above expressions usually gives a reasonable 
approximation of $c_{i\ell}$
for small values of $\ell$ as well.
If this is the case, we
can evaluate the infinite product and derive an approximate expression for
the prefactors\Blue
\be
\ln A_{ik} \approx {\rm sign}(D_i) \sqrt{|D_i|} \pi.\Black
\label{apprpref} \ee
We assume that $A_{ik}$ are measured in units of some dimensionful quantity of
the order of the scale $k_f$. The precise determination of these
units is not necessary because of the approximate nature of the above
expression.

\begin{figure*}[t]
\begin{center}\hspace{-5mm}
\begin{picture}(17,11)
\putps(0,0)(-0.5,0){fSAapprox}{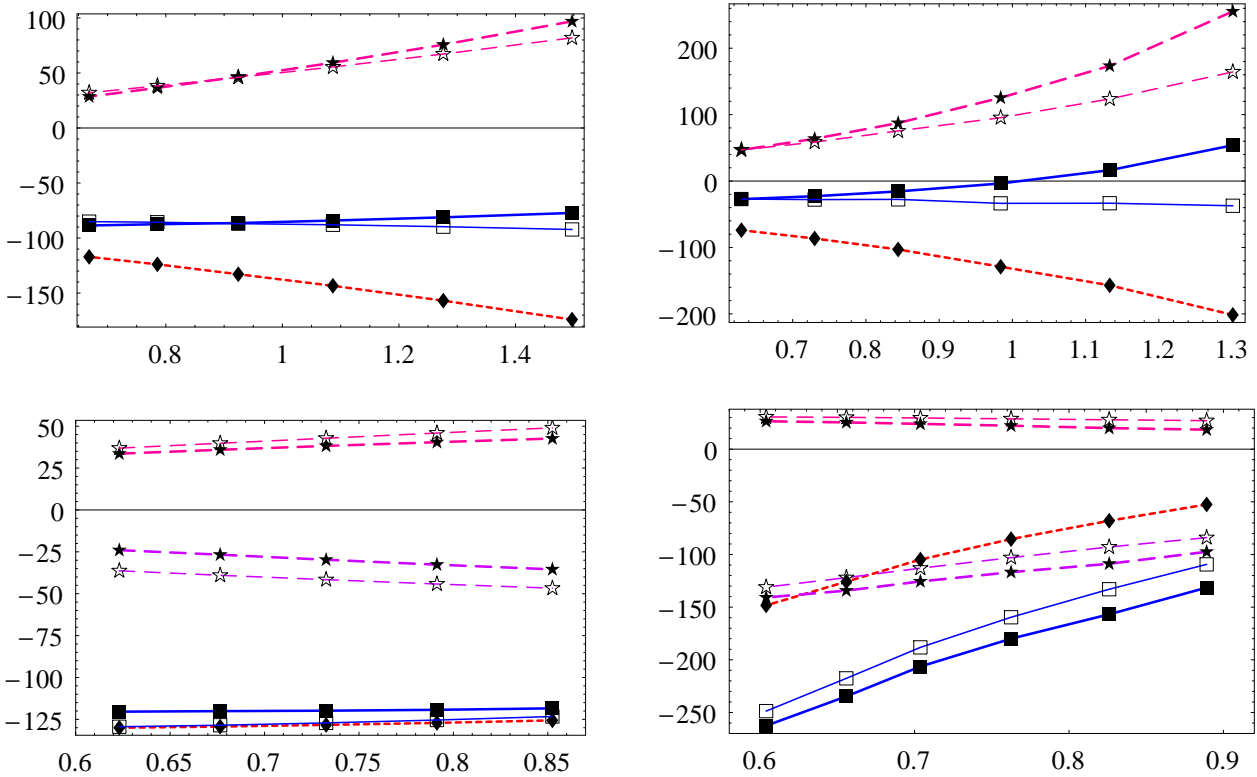}
\end{picture}
\caption[SP]{\em 
The negative of the action of the saddle point 
$-S_k$ (diamonds), the 
prefactor $\ln ( A_k/k^4_f )$
(stars) and the nucleation rate  $\ln( I/k^4_f )$
(squares), as a function of the scale $k$ in units of the field mass
at the absolute minimum.
The dark stars and squares denote values computed numerically, 
while the open stars and squares denote
values obtained through the estimate of eq.~$(\ref{apprpref})$. 
The first row corresponds to a theory of one scalar field, and
the second row to a two-field theory.
\label{fig:approx}}
\end{center}\end{figure*}

The validity of eq.~(\ref{apprpref})
can be checked through a comparison
with the numerical evaluation of the prefactors in refs.~\cite{first}--\cite{third}.
In fig.~1 we present four examples from these studies, in
which we have added the predictions of eq.~(\ref{apprpref}). Figs.~1a and 1b
correspond to the one-scalar theory of refs.~\cite{first,second}, 
for the same choice of parameters as in fig.~1 of ref.~\cite{second}.  
The initial potential at the scale $k_0$ has already two minima separated by
a barrier. The form of the potential at lower scales is determined through
the numerical integration of eq.~(\ref{evpot}), with $\Mc^2_k(\phi)=
U''_k(\phi)$. The negative of the action of the saddle point 
$-S_k$ is denoted by diamonds, the 
prefactor $\ln ( A_k/k^4_f )$
by stars and the nucleation rate  $\ln( I/k^4_f )$
by squares. The dark stars and squares denote values computed numerically, 
while the open stars and squares denote
values obtained through the estimate of eq.~(\ref{apprpref}). The above
quantities are plotted as a function of the ratio
$k/\sqrt{U''_k(\phi_t)}$, i.e. the scale $k$ in units of the field mass
at the absolute minimum located at $\phi_t$. 
In the first case the expansion around the saddle point is convergent, while
in the second one it breaks down. For both cases, we observe a good 
agreement between the exact and the approximate values 
of $\ln ( A_k/k^4_f )$
for the range of
scales near $k_f$. 

In figs.~1c and 1d we present similar results for 
the two-scalar theory of ref.~\cite{third}.
There are two prefactors, 
corresponding to the field $\pha$ that varies along the profile of the bubble, 
and the
orthogonal field $\phb$ with zero expectation value.
In the case of fig.~1c (corresponding to fig.~2 of 
ref.~\cite{third}), the
potential at the scale $k_0$ has already two minima. In the case of 
fig.~1d (corresponding to fig.~4 of 
ref.~\cite{third}),
the 
initial potential is
$U_{k_0}=m^2_{k_0}(\pha^2+\phb^2)/2
+\lambda_{k_0}(\pha^4+\phb^4)/4
+g_{k_0}\pha^2 \phb^2/4$ with $g_{k_0}/\lambda_{k_0}=40$\footnote{
There is a difference of a factor of 2 in the definition of $\lx$ 
between this work and ref.~\cite{third}.}.
The minimum at the origin appears at scales below $k_0$ and the 
first-order phase transition is radiatively induced~\cite{colwein}. 
In the first case the expansion around the saddle point is convergent, while
in the second one, which is typical of all radiatively-induced phase 
transitions we have studied, it breaks down. 
For both cases, we observe again a good 
agreement between the exact and the approximate values of 
the prefactor.

\paragraph{Metastability bound on the Higgs-boson mass:}
As a first example we apply our formalism to the question of the 
bound on the Higgs-boson mass from the stability of our vacuum
\cite{stab}--\cite{jose}. The top-quark radiative corrections to the effective
potential of the Higgs field may result in the appearance of a new
minimum, deeper than the one located at 247 GeV. In order for this not to
happen, a lower bound on the Higgs-boson mass must be imposed~\cite{stab}.
This bound can be relaxed if one allows for the presence of a new vacuum,
but demands that the time needed for a transition to
it is larger than the age of our universe. The largest 
rates are associated with thermal transitions at high temperatures\footnote{
The typical temperature at which the nucleation rate is maximized is of
the order of the Higgs-field value at the absolute minimum. This can
be larger than 247 GeV by several orders of magnitude. It is implicitly
assumed that such temperatures were realized in the early Universe, for example
through efficient reheating after inflation. If this is not the case, the 
bound on the Higgs-boson mass is less stringent than the one we consider.
}, from
a metastable minimum in the symmetric phase of the Standard Model directly to 
the absolute minimum at very large Higgs-field expectation values 
\cite{anderson}--\cite{jose}. 

In our simplified
discussion we neglect the contributions from the gauge fields to the
effective potential of the Higgs field. The main effect of interest
is associated with the top quark, while our estimates of the transition rate
are approximate.
As a result, a detailed discussion of the gauge-field contributions
does not improve the accuracy of our conclusions. 
We can start by considering the effective three-dimensional description of
the top-Higgs system at a scale $k_0=T$. At this scale the top-quark 
fluctuations are expected to be almost decoupled, as the absence of a
zero Matsubara frequency for fermions implies.
This permits a very simple determination of the potential $U_{k_0}(\phi)$, 
which is equal to the sum of the zero-temperature potential and
the contribution from the top-quark thermal fluctuations. The latter can be
approximated by the standard one-loop expression.
There is also the contribution of the Higgs-boson thermal fluctuations 
with characteristic momenta $q^2 > k^2_0$~\cite{trans,me}, 
but their effect is small for the Higgs-boson masses of interest.

Following ref.~\cite{arnold}, we keep only the leading temperature
correction in the expansion of the one-loop thermal
contribution from the top quark. By performing the rescaling of
eqs.~(\ref{fivethree}), we express the potential as
\be
U_{k_0}(\phi) = \frac{m^2}{2} \phi^2 - \frac{\kx}{4}  \phi^4,
\label{toppot} \ee
where
$m^2 = g^2_{t4}\, T^2/4$, 
$\kx = |\lx_4| \, T$, 
and $\lx_4$ is the average (negative) value of the quartic coupling of
the Higgs field over the region of Higgs-field expectation values
relevant for the problem\footnote{
As the zero-temperature quartic coupling $\lx_4$ varies only logarithmically
with $\phi$, it can be approximated as constant over the range of 
field values that are relevant for the profile of the critical bubble
\cite{arnold}.}.
The absolute minimum of the potential lies at Higgs-field values 
larger than the range over which the above approximation is valid. 
However, the profile of a critical bubble and the nucleation rate can be
determined, exactly as in ref.~\cite{arnold}. 
For the determination of the 
prefactor we need to 
specify the coarse-graining scale $k_f$.
As the potential at the scale $k_0=T$ already has a metastable minimum
and a barrier, the choice $k_f=k_0$ seems justified. 

Through the rescalings
$r=\rt/m$, $\phi=\phit m/\sqrt{\kx}$, the potential
can be written as 
$\Vt(\phit)=\phit^2/2-\phit^4/4$. As a result, we need to
determine
the saddle-point profile numerically only once. The action of the 
saddle point can be written as~\cite{arnold}
\be
S_{k_0} =  \St \,\frac{m}{\kx}  = 18.9 \frac{m}{\kx} = 9.45 \, 
\frac{g_{t4}}{|\lx_4|}.
\label{action1} \ee
Using the
approximate expression~(\ref{apprpref}), we find 
\be
\ln  A_{k_0}  \approx 2.00 \, \frac{k_0 \, \pi}{2 \, m}=
\frac{6.28}{g_{t4}}.
 \label{prefactor1} \ee

The general form of eqs.~(\ref{action1}), (\ref{prefactor1}) does not
guarantee that the prefactor is smaller than the saddle-point action
for all values of $g_{t4}$. Our estimate 
for the prefactor grows in inverse proportion to the Yukawa coupling 
and can be significant for small $g_{t4}$. 
The calculation of $\lx_4$ 
is necessary for the determination of the saddle-point
action. 
For the experimental value of the top-quark mass ($g_{t4} \approx 1$) we find
$|\lx_4|\lta 0.05$ for all Higgs-field expectation values, 
in agreement with fig.~1 of ref.~\cite{arnold}. 
As a result, the prefactor is expected to give only a small correction to the 
bubble-nucleation rate. We mention at this point that, for the experimental
value of the top-quark mass ($m_t=174$ GeV)~\cite{top}
and the experimental lower bound 
on the Higgs-boson mass ($m_H > 90$ GeV) 
\cite{higgs}, the difference between the bound 
from absolute vacuum stability and the more relaxed
metastability bound we discussed is small~\cite{jose}. 
This is an additional 
reason that makes a more refined determination of the prefactor unnecessary.

\paragraph{The electroweak phase transition:}
We turn next to the case of radiatively-induced first-order phase
transitions. The potential at the scale $k_0=T$ has only one minimum,
while a new one is generated at a lower scale.
Several examples have
been given in the past of numerical or analytical 
solutions of evolution equations, such as eq.~(\ref{evpot}),
that display this behaviour~\cite{twoscalar,third}.
The basic picture can be obtained by considering the 
approximate solution of eq.~(\ref{iter}) for a theory of two scalar fields
(for simplicity we assume the symmetries $\pha \leftrightarrow - \pha$, 
$\phb \leftrightarrow - \phb$, $\pha \leftrightarrow \phb$).
The expectation value of $\pha$ is the order parameter
for the phase transition, while we consider an
interaction $g_4 \pha^2 \phb^2/4$ between the two fields.
The contribution $\propto[ k_0^2 + \Mc^2_{k_1}]^{{3}/{2}}$
in eq.~(\ref{iter})
can be expanded in powers of $\Mc^2_{k_1}/k_0^2$.
The first term in this expansion, when added to the 
temperature-dependent part in $U_{k_0}$, reproduces the
leading high-temperature behaviour of the one-loop perturbative
effective potential, as has been checked
explicitly in refs.~\cite{trans,me,twoscalar}.
Thus, we can write
\beq
U_k^{(1)}(\pha) \approx -\frac{1}{2} \mu^2 \pha^2 + \frac{1}{4} \lx \pha^4
+ \frac{1}{48} (6\lx + g ) k_0 \pha^2  \label{pot1}
\eeq
$$
-\frac{1}{12\pi} \left\{
 [ k^2 + \Mc^2_{1k_1}(\pha) ]^{\frac{3}{2}} 
+[ k^2 + \Mc^2_{2k_1}(\pha) ]^{\frac{3}{2}}
\right\},$$
with $\lx=\lx_4 T$, $g= g_4 T$.
Setting $k=k_1=0$ in the above equation results in the standard
``mass-gap'' equation for the high-temperature potential of this model.
As we explained earlier, the calculation of the nucleation rate
must be performed for a non-zero value $k_f$, with
$k^2_f = \max |U_{k_1}''{}^{(1)}(\phi)|$. 
For this reason, we
use $k=k_1=k_f$ in the following.
For radiative symmetry breaking, 
$g$ is in general much larger than $\lx$. 

For an approximate estimate of the nucleation rate we can make
certain simplifications in eq.~(\ref{pot1}).
We can approximate the mass term of the $\phb$ field by
its zero temperature form: $\Mc^2_{2k}(\pha) \approx g \pha^2 /2$. An improved
treatment would take into account the thermal corrections to this mass and
the $k$ dependence of the three-dimensional coupling $g$. However, both the
above corrections are expected to be small for strongly first-order phase
transitions, on which we focus for the rest of our discussion.
Moreover, for $g < 1$ and $k=k_1=k_f$, 
we can neglect $k^2$ relative to $\Mc^2_{2k_1}(\pha)$. 
This approximation is not valid 
only for a small region around the origin, while most
of the potential
(and especially the barrier) 
remains unaffected. The term 
$\propto [ k^2 + \Mc^2_{1k_1}(\pha)]^{{3}/{2}}$ 
in eq.~(\ref{pot1}) 
can also be neglected, as it
is associated with fluctuations of the $\pha$ field whose
self-coupling $\lx$ is much smaller than $g$.
We conclude that,
for an estimate of the bubble-nucleation rate for strongly first-order 
phase transitions, we can use
\be
U_{k_f}^{(1)}(\pha) \approx
\frac{1}{2} m^2  \pha^2 
-\frac{1}{3} \gamma \pha^3
+ \frac{1}{4} \lx \pha^4,
\label{pot3} \ee
with
\be
m^2 = \frac{g k_0}{24}    -\mu^2,\quad\quad
\gamma = \frac{g^{\frac{3}{2}}}{8\sqrt{2}\pi} ,\quad\quad
k^2_f = \frac{\gamma^2}{3\lx}-m^2.
\label{abc} \ee
The potential of eq.~(\ref{pot3})
is nothing but the high-tempe\-rature expansion of the
one-loop approximation, for which only the dominant
$\phb$ fluctuations have been considered. The crucial additional
piece of information that has been obtained through our discussion
is the value of the coarse-graining scale $k_f$ at which 
this potential can be used for the calculation of the
bubble-nucleation rate. 

\begin{figure}[t]
\begin{center}\hspace{-5mm}
\begin{picture}(9,5)
\putps(0,0)(0,0){fRh}{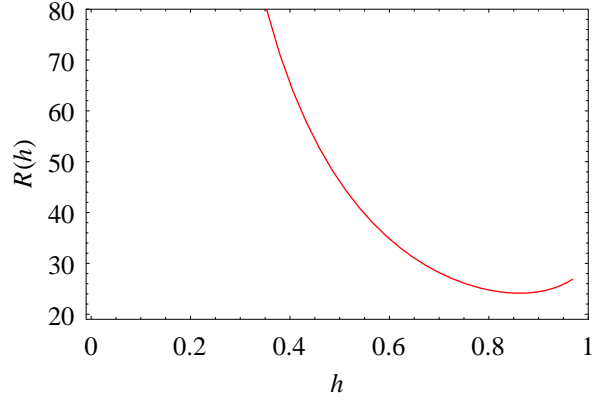}
\end{picture}
\caption[SP]{\em The parameter $R(h)$, defined in eq.~$(\ref{fin})$,
as a function of $h.$
\label{fig:SAh}}
\end{center}\end{figure}

Through the rescalings 
$r=\rt/m$, $\pha=\phit_1\, m^2/\gamma$, the potential
can be written as 
$\Vt(\phit_1)=\phit_1^2/2-\phit_1^3/3+h\,\phit_1^4/18$, with
$h=9\lx m^2/2\gamma^2$. 
For $h \approx 1$ 
the two minima of the potential have approximately
equal depth. The action of the saddle point can be expressed
as
\be
S_{k_f} = \frac{1}{108\,\pi}
\left( \frac{g h}{\lx}\right)^{\frac{3}{2}}\St(h),
\label{action2} \ee
where $\St(h)$ must be determined numerically through $\Vt(\phit_1)$.
Similarly, the pre-exponential factor associated with the $\phb$ field
can be estimated from eq.~(\ref{apprpref}) as 
\beq
\ln  A_{2k_f}  &\approx& 
-\frac{\pi}{6 \sqrt{2}}
\left[ \frac{g( 3 -2 h)}{\lx}\right]^{\frac{1}{2}} \At(h),
\nonumber \\ 
\At^2(h) &=& 
\int_0^\infty \phit^2_1(\rt) \,\rt^3\, d\rt,
\label{prefactor2} \eeq
with $\At(h)$ computed numerically.
Finally
\be
\frac{\ln  A_{2k_f} }{S_{k_f}} \approx
-9\sqrt{2} \pi^2 \, \frac{
( 3 -2 h)^{\frac{1}{2}}
\At(h)}{h^{\frac{3}{2}}\St(h)} \, \frac{\lx}{g}=
- R(h) \,\frac{\lx}{g}.
\label{fin} \ee
In fig.~2 we plot $R(h)$ as a function 
of $h$ in the interval (0,\,1). This function has 
a minimum $R_{\rm min}\approx 24$.
In the limit $h \to 0$, our estimate of the prefactor
predicts a constant value for
$R(h)$. The reason is that 
the radius $\Rt$ 
of the critical bubble becomes very large in this limit, while 
$\At^2(h) \propto\Rt^4$, $\St(h) \propto\Rt^2$. 
However, the approximate expression~(\ref{apprpref}) has not
been tested for very large critical bubbles. The divergence of the 
saddle-point action and the prefactor in this limit results in low accuracy
for our numerical analysis. Typically, our results are reliable
for $\St(h)$ less than a few thousand.
On the other hand, eq.~(\ref{apprpref})
relies on the large-$\ell$ approximation of 
eq.~(\ref{asympt}). For increasing $D_i$ this approximation
breaks down below an increasing value $\ell_{as}$ and, therefore, 
eq.~(\ref{apprpref}) is not guaranteed to be valid.
However, for the strongly first-order phase transitions that we
are discussing, $g$ is larger than $\lx$ by at least one order of magnitude
and values $S_{k_f}={\cal O}(100)$ in 
eq.~(\ref{action2}) can be obtained for $h \lta 0.9$, where
both our numerical and approximate results are reliable. 

From eq.~(\ref{fin}) and fig.~2 we conclude that 
$$|{\ln  A_{2k_f} }|/{S_{k_f}} \gta 
24\,\lx/g.$$
This result is in agreement with fig.~1d, for which
$g/\lx=40$ and $S_{k_f}$, $|{\ln  A_{2k_f}}|$
are comparable. 
Also, the increase of the above ratio for decreasing 
$g/\lx$ is in quantitative agreement with the results of ref.~\cite{third}.
More specifically, the predicted doubling of 
$|{\ln  A_{2k_f} }|/{S_{k_f}}$ for 
$g/\lx=20$ is confirmed by fig.~5 of that study.
We conclude that the expansion around the saddle point is not convergent,
unless $\lx$ is smaller than $g$ by at least two orders of magnitude.
In the limit of zero $\lx$ (the Coleman-Weinberg limit~\cite{colwein})
our analysis is not sufficient, as corrections $\propto \pha^4 \ln \pha$
that we have neglected become important, and 
eq.~(\ref{pot3}) is not a good approximation any more. 
However, this limit is not relevant for the electroweak phase 
transition, which is the application we have in mind.

The behaviour of the prefactor associated with the $\phb$ field
can be explained by the form of the differential operators in it.
$A_{2k_f}$
involves the ratio 
$$\det\big[-\partial^2+\Mc^2_{2k_f}(\phi_b(r))\big]/
\det\big[-\partial^2+\Mc^2_{2k_f}(0)\big],$$ with
$\Mc^2_{2k_f}(\pha)\approx g\,\pha^2/2$.
It is easy to check from 
eq.~(\ref{pot1}) that, unless $k^2_f \ll \Mc^2_{2k_f}(\pha)$ 
for all values of $\pha$ apart from
a small region around the origin of the potential, the two-minimum 
structure cannot be generated. 
This implies that, in 
units of $k_f$, 
the function $\Mc^2_{2k_f}(\phi_b(r))$ takes very large positive values
near $r=0$ (see figs 4, 5 of ref.~\cite{third}).
As a result, the lowest
eigenvalues of the operator $\det[
-\partial^2+\Mc^2_{2k_f}(\phi_b(r))]$ are 
much larger 
than those of $\det[-\partial^2+\Mc^2_{2k_f}(0)]$. 
This induces a large suppression of the nucleation rate. 
In physical terms, it implies that the deformations of the critical
bubble in the $\phb$ direction cost excessive amounts of free energy, because
the $\phb$ field is very massive apart from near the origin of the potential. 
As the $\phb$ fluctuations are inherent to the system, the total nucleation 
rate is suppressed considerably when their effect is taken into account.

Based on the above discussion, we can estimate the importance of
the pre-exponential factor of the bubble-nucleation rate for
the electroweak phase transition. We should point out that the 
rigorous implementation of our approach in gauged Higgs systems must
deal with several difficult issues, such as the gauge-invariant 
definition of the nucleation rate, the gauge-invariant implementation
of an ultraviolet cutoff in the prefactor, and the strongly-coupled
symmetric phase. However, the basic picture 
is expected to be very similar to the one in the simple model 
we considered, at least for strongly first-order phase transitions
for which the detailed treatment of the symmetric phase is not
crucial. 
One can use again an effective three-dimensional description,
because the longitudinal
gauge-field fluctuations develop a large thermal mass and 
can be integrated out.

The form of the potential is expected to be given by 
an expression similar to eq.~(\ref{pot3}).
The parameter 
$\gamma$ includes now contributions $\propto g^3_i$
from fluctuation determinants
(around a constant field configuration)
involving fields with couplings $g_i$ to the Higgs field. 
Analogous fluctuation determinants must be 
included in the prefactor, which becomes a product of terms similar to 
the one of eq.~(\ref{prefactor2})\footnote{
We approximate the fluctuation determinant by a product of
determinants, assuming a diagonal mass matrix. Off-diagonal
elements, such as the ones appearing in the Goldstone-boson sector
in background gauges 
\cite{schmidt} are $\propto d\phi_b/dr$ and, therefore,
smaller than the diagonal elements 
$\propto \phi_b^2$ for a slowly varying critical-bubble background.}. 
Each of these
terms is $\propto g_i$. The final estimate of 
the magnitude of the prefactor is given by eq.~(\ref{fin}), with
$\lx/g$ taking the ``effective'' value
\be
\left(\frac{\lx}{g}\right)_{\!\!\rm SM} \approx 
\frac{(4 m_W+2 m_Z)\,m^2_H}{
4( 4 m^3_W+2 m^3_Z)}=0.35 \, \left( \frac{m_H}{{\rm 100 \,GeV}}\right)^2.
\label{finn} \ee
For the prefactor $|{\ln  A_{2k_f} }|$, associated
now with the gauge-field sector, to be smaller than
1/2 of the action\footnote{
As a very mild criterion of the reliability of the expansion around the saddle
point, we demand that the first-order correction (the prefactor) is smaller
than 1/2 of the saddle-point action. The bound on the Higgs-boson mass for
alternative choices can be obtained easily from 
eqs.~(\ref{fin}), (\ref{finn}).}, 
the Higgs boson mass must be below $\sim 25$ GeV. Our conclusions about the 
large suppression of the nucleation rate by the pre-exponential factor
associated with the gauge-field fluctuations is consistent with the
results of refs.~\cite{schmidt}.

Of course, Higgs boson masses below the experimental lower limit of 90 GeV 
\cite{higgs} are of academic interest. Moreover, 
for $m_H \gta 80$ GeV, there is
no phase transition in the high-temperature Standard Model
\cite{crossover}.
There is the possibility, however, for a sufficiently stron\-gly 
first-order phase transition for baryogenesis within the Minimal
Supersymmetric Standard Model in presence of a very 
light stop~\cite{mssm}. 
In this model, the ratio of the ``effective'' $g$ to $\lx$ obeys\footnote{
The fermionic partners of the various fields do not have a zero  
Matsubara frequency and 
are not relevant for the effective three-dimensional theory. Moreover,
they do not affect the term $\propto \pha^3$ of the potential.}
\begin{eqnarray}
\left(\frac{\lx}{g}\right)_{\!\!\rm MSSM} &\gta &
\frac{(4 m_W+2 m_Z+6 m_t)\,m^2_H}{
4( 4 m^3_W+2 m^3_Z+6 m^3_t)} \nonumber\\
&=&0.11 \, \left( \frac{m_H}{{\rm 100 \,GeV}}\right)^2.
\label{finnn} 
\end{eqnarray}
For $m_t= 175$ GeV~\cite{top},
in order to have
$|{\ln A_{2k_f} }|$ smaller than
1/2 of the action, the Higgs boson mass must be below $\sim 45$ GeV. 

It must be pointed out that the additional suppression of the 
nucleation rate by the prefactor is beneficial for baryogenesis,
as it delays the phase transition. This results in larger 
Higgs expectation values in the interior of the critical bubbles
that 
suppress sphaleron transitions. However, the possible
non-convergence of
the expansion around the saddle point prohibits any quantitative
predictions for the true value of the nucleation rate. 
Another important point concerns the prefactor associated with
the Higgs field fluctuations (the $\pha$ field in our discussion).
For weakly first-order phase transitions, such as the ones for
Higgs-boson masses larger than the masses of the gauge bosons, 
this prefactor can 
enhance the total rate considerably and even compensate the
exponential suppression~\cite{first,second}.
However, these large corrections 
signal the breakdown of the saddle-point approximation.

The presence of significant higher-order corrections
in the expansion around the saddle point 
is confirmed by a large residual dependence of the predicted nucleation
rate on the ``renormalization'' scale $k$.
On the contrary, this dependence  
is very small when the expansion is convergent. This behaviour has been
demonstrated numerically 
in refs.~\cite{first}--\cite{fourth} and was not discussed
in this paper. Instead we focused here on the presentation of simple
analytical expressions (eqs.~(\ref{iter}), (\ref{asympt}), (\ref{apprpref})) 
that can be used for fast checks
of the reliability of the estimated nucleation rates for cosmological phase
transitions.

\paragraph{Acknowledgements:} We  would like to thank
R. Barbieri, J. Espinosa and C. Wetterich for helpful discussions.
The work of N.T. was supported by the E.C. under TMR contract 
No. ERBFMRX--CT96--0090.

\end{document}